\documentclass{article}
\usepackage{spconf,amsmath,graphicx}
\usepackage {setspace}
\usepackage{booktabs}
\usepackage{fancyhdr}
\usepackage{tabularx}
\usepackage{caption} 

\title{VARIATIONAL BAYESIAN APPROXIMATIONS KALMAN FILTER BASED ON THRESHOLD JUDGMENT}
\twoauthors
  {Zuxuan Zhang, Gang Wang*}
	{School of Information and\\ Communication Engineering\\
	University of Electronic Science and\\ Technology of China\\
	Chengdu P.R. China 611731}
  {Jiacheng He, Shan Zhong}
	{School of Mechanical and\\ Electrical Engineering\\
	University of Electronic Science and\\ Technology of China\\
	Chengdu P.R. China 611731}
\begin{document}
%
\maketitle
\thispagestyle{fancy}
\lhead{}
\lfoot{}
\cfoot{\small{\copyright~2023 IEEE.Personal use of this material is permitted. Permission from IEEE must be obtained for all other uses, in any current or future media, including reprinting/republishing this material for advertising or promotional purposes, creating new collective works, for resale or redistribution to servers or lists, or reuse of any copyrighted component of this work in other works}}
\rfoot{}
\begin{abstract}
	
	The estimation of non-Gaussian measurement noise models is a significant challenge across various fields. In practical applications, it often faces challenges due to the large number of parameters and high computational complexity. This paper proposes a threshold-based Kalman filtering approach for online estimation of noise parameters in non-Gaussian measurement noise models. This method uses a certain amount of sample data to infer the variance threshold of observation parameters and employs variational Bayesian estimation to obtain corresponding noise variance estimates, enabling subsequent iterations of the Kalman filtering algorithm. Finally, we evaluate the performance of this algorithm through simulation experiments, demonstrating its accurate and effective estimation of state and noise parameters.
\end{abstract}
\section{Introduction}
\label{sec:intro}

Kalman filter\cite{1960A} is a widely used algorithm for state estimation in various fields, known for its effectiveness in handling Gaussian noise and known noise variances. It has found applications in diverse domains, including navigation, control systems, and signal processing. Despite its success, the performance of traditional Kalman filtering algorithms deteriorates when faced with practical scenarios where the observation noise characteristics are uncertain\cite{1999Non}. Therefore, there is a need to develop robust algorithms that can handle unknown noise variances and non-Gaussian noise distributions to improve state estimation accuracy.

In practice, it is difficult to obtain the variance of observed noise. The uncertainty of noise variance significantly affects the performance of traditional Kalman filtering algorithms. In order to solve the problem of unknown noise variance, a method called Variational Bayesian Adaptive Kalman Filtering (VBAKF) is proposed in recent years\cite{2003Variational,2009Recursive}. This method uses variational Bayesian approximation recursively to estimate unknown noise variance. The algorithm has a limitation that it cannot deal with non-Gaussian noise due to the inherent Gaussian noise distribution assumption in variational Bayesian frame\cite{KFn}.

The observation noise is non Gaussian noise is often encountered in practical applications, and its variance amplitude change has a significant impact on the effectiveness of the Kalman filter\cite{2010Bridging}. In the BIKF algorithm\cite{4}, Fan utilizes the EM algorithm\cite{1977Maximum} to address the challenge of non-Gaussian noise by estimating the parameters of Gaussian mixture models. In addition to the BIKF algorithm, another approach that addresses non-Gaussian noise is the Maximum Entropy Kalman Filter (MCKF), which is based on the maximum correlation entropy criterion\cite{2017Maximum}. However, similar to the BIKF algorithm, both the MCKF and BIKF algorithms face certain limitations\cite{2,3}.

We proposed the T-VBKF(Variational Bayesian Approaches Kalman Filter), designed to address challenges related to unknown observation noise variances and non-Gaussian noise\cite{nGS}. It employs a threshold judgment approach to classify iteration point noise and utilizes variational Bayesian methods for Gaussian mixture model parameter inference. By incorporating adaptive and non-Gaussian noise handling, T-VBKF surpasses current methods, delivering more precise state estimation in practical applications\cite{1}.

\section{PROPOSED T-VBKF}
\subsection{Kalman Filter}
In linear systems, the Kalman filtering algorithm effectively solves the problem of state estimation under Gaussian noise.
Consider a linear system described by the following state and measurement equations:
\begin{align}
	{x}_{k} &= {A}_{k-1}{x}_{k-1} + {q}_{k-1} \\
	{y}_{k} &= {H}_{k}{x}_{k} + {r}_{k}
\end{align}
where $q$ is the Gaussian process noise, $r$ is the measurement noise with diagonal covariance. The measurement $x_k$ is a n-dimensional vector and the state is an n-dimensional vector.

We can summarize the KF algorithm as follows.
\begin{align}
	\hat x_k^- &= A\hat x_{k-1} \\
	{P}_{k|k-1} &= A_k{P}_{k-1}A_k^T + Q_k \\
	K_k &= {P}_{k|k-1}H_k^T\left(H_k{P}_{k|k-1}H_k^T + \hat R_k^{(n)}\right)^{-1} \\
	\hat{x}_{k}^{(n+1)} &= \hat{x}_{k|k-1} + K_{k}\left( y_{k} - H_{k}\hat{x}_{k|k-1} \right) \\
	P_{k}^{(n+1)} &= {P}_{k|k-1} - K_{k}H_{k}{P}_{k|k-1}
\end{align}

\subsection{Variational Bayesian Kalman Filter}
Variational Bayesian Adaptive Kalman Filtering Algorithm (VBAKF) is introduced to deal with uncertainty and noise models.The key idea of this method is to treat the system state and observation data as probability distributions and update them by Bayesian inference.

The core idea of the algorithm is variational inference. In this stage, we use the variational inference method to find an approximate posterior distribution to replace the true posterior distribution. This approximate distribution can be represented by a simple distribution, such as a Gaussian distribution or an exponential distribution. 
\begin{equation}
	p\left( {{x_k},{\Sigma _k}\mid {y_{1:k}}} \right) \approx {Q_x}\left( {{x_k}} \right){Q_\Sigma }\left( {{\Sigma _k}} \right)
\end{equation}

By minimizing the Kullback Leibler divergence, we can obtain an optimal approximate posterior distribution.
\begin{equation}
	\begin{array}{l}
		{\rm{KL}}\left[ {{Q_x}\left( {{x_k}} \right){Q_\Sigma }\left( {{\Sigma _k}} \right)p\left( {{x_k},{\Sigma _k}\mid {y_{1:k}}} \right)} \right] = \\
		\int {{Q_x}} \left( {{x_k}} \right){Q_\Sigma }\left( {{\Sigma _k}} \right) \times \log \left( {\frac{{{Q_x}\left( {{x_k}} \right){Q_\Sigma }\left( {{\Sigma _k}} \right)}}{{p\left( {{x_k},{\Sigma _k}\mid {y_{1:k}}} \right)}}} \right){\rm{d}}{x_k}\;{\rm{d}}{\Sigma _k}
	\end{array}
\end{equation}

\subsection{Threshold Judgment Criteria}
\subsubsection{Non-Gaussian Measurement Noises}

Non Gaussian noise refers to noise that does not conform to a Gaussian distribution. In signal processing, non Gaussian noise is often considered a more realistic noise model.
\begin{figure}
	\centering
	\includegraphics[scale=0.7]{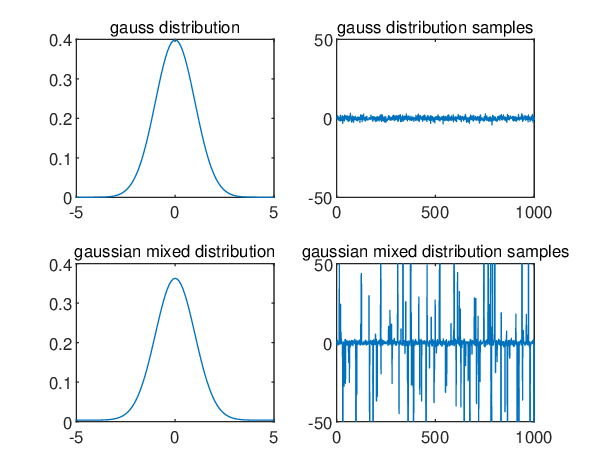}
	\captionsetup{skip=0pt} 
	\caption{Noise Samples}
	\label{1}
\end{figure}

As shown in the Fig. \ref{1}, in a group of Gaussian noises, the sample data distribution is relatively concentrated. When the noise is Gaussian mixture model, more "outlier" with large differences from the mean will appear\cite{ye2009efficient}.
\subsubsection{Akaike information criterion}
Akaike Information Criterion (AIC) can be used to estimate the fitting superiority of the model. When several sets of models are provided for the same data set, the AIC criterion can evaluate the fit of the model relative to the set of data. Therefore, AIC provides a model selection method\cite{AIC}.
\begin{equation}
	{\rm{AIC}} = 2k - 2\ln (\hat L)
\end{equation}

Given several candidate models for a set of data, the optimal model is the one with the lowest AIC value.AIC evaluates the goodness of fit by likelihood function, and also includes penalty terms for increasing the number of estimated parameters.

The likelihood function of GMM is the product of the likelihood of two different normal distributions, so it has four parameters: ${\mu_1}$, ${\sigma _1}$, ${\mu _2}$, ${\sigma _2}$. To be explicit, the likelihood function is as follows (denoting the sample sizes by $N_{1}\left( {{\mu }_{1}},\sigma _{1}^{2} \right)$ and $N_{2}\left( {{\mu }_{2}},\sigma _{2}^{2} \right)$).
\begin{equation}
	\begin{split}
		{\cal L}({\mu _1},{\sigma _1},{\mu _2},{\sigma _2}) = 
		\prod\limits_{i = 1}^{{n_1}} {\frac{1}{{\sqrt {2\pi } {\sigma _1}}}} \exp \left( { - \frac{{{{({x_i} - {\mu _1})}^2}}}{{2\sigma _1^2}}} \right) \\
		\cdot \prod\limits_{i = {n_1} + 1}^{{n_1} + {n_2}} {\frac{1}{{\sqrt {2\pi } {\sigma _2}}}} \exp \left( { - \frac{{{{({x_i} - {\mu _2})}^2}}}{{2\sigma _2^2}}} \right)
	\end{split}
\end{equation}

In these two sets of mixed Gaussian distribution models, we assume that the mean of the two Gaussian mixture distributions is 0, and we do not know the mixing proportion and variance of the two Gaussian distributions.

We divide the Gaussian mixture model into two Gaussian distributions, where k=1.
\begin{equation}
	\begin{split}
		{\rm{AI}}{{\rm{C}}_{{\sigma _1},{\sigma _2}}} = 2k - 2\ln (\hat L({\mu _1},{\sigma _1},{\mu _2},{\sigma _2}))\\
		= 2 - 2\ln (\prod\limits_{i = 1}^{{n_1}} {\frac{1}{{\sqrt {2\pi } {\sigma _1}}}} \exp \left( { - \frac{{{{({x_i} - {\mu _1})}^2}}}{{2\sigma _1^2}}} \right) \\
		\cdot \prod\limits_{i = {n_1} + 1}^{{n_1} + {n_2}} {\frac{1}{{\sqrt {2\pi } {\sigma _2}}}} \exp \left( { - \frac{{{{({x_i} - {\mu _2})}^2}}}{{2\sigma _2^2}}} \right))
	\end{split}
\end{equation}

\subsubsection{Calculate Threshold}
\begin{figure}[htb]
	\begin{minipage}[b]{1.0\linewidth}
		\centering
		\includegraphics[scale=0.13]{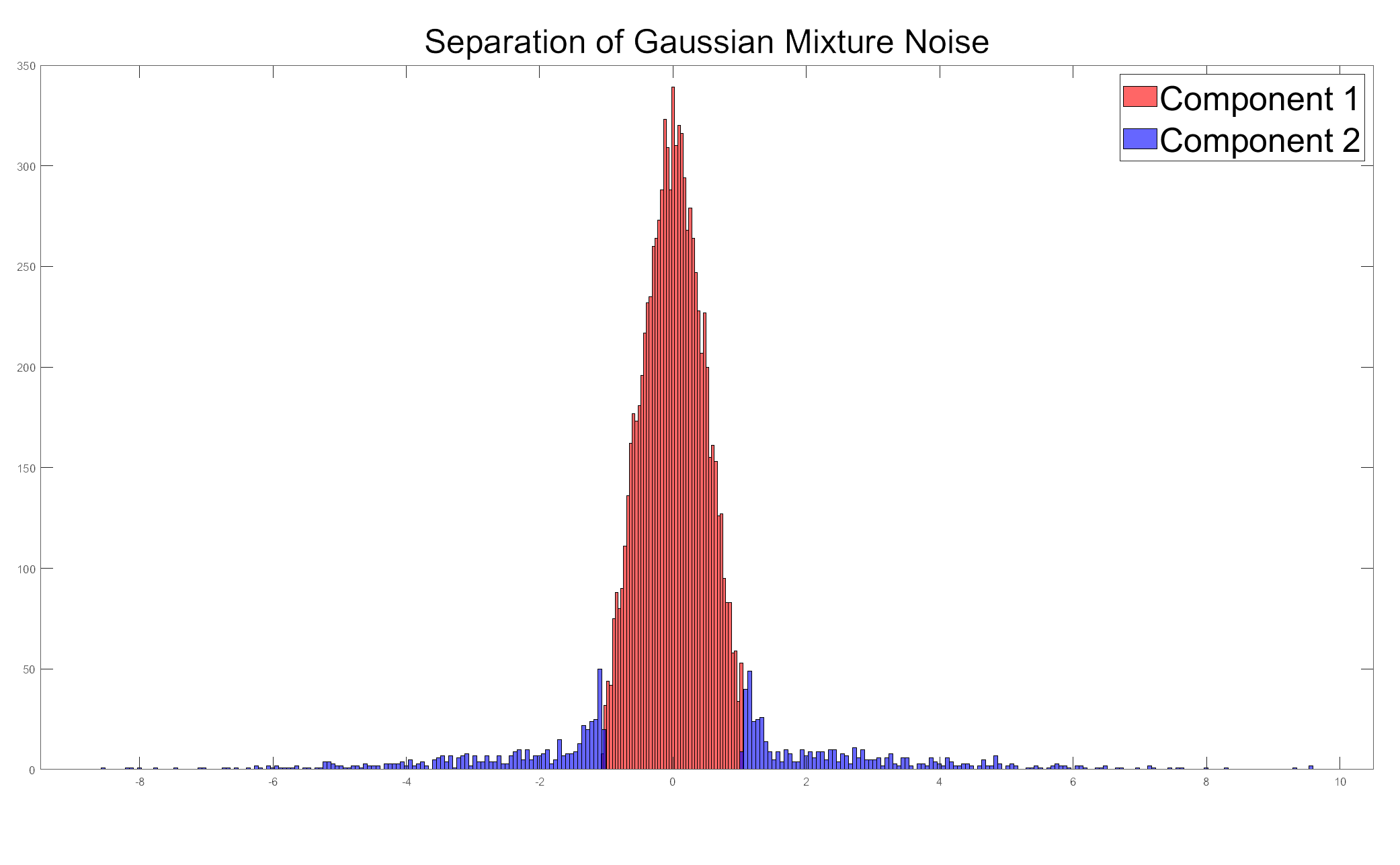}
	\end{minipage}
	\captionsetup{skip=0pt} 
	\caption{Separation of Gaussian Mixture Noise}
	\label{2}
\end{figure}

In a normal distribution, the distribution of data follows the 3-$\sigma$ rule rule. Here, we attempt to consider using $\sigma$ as a threshold to divide a set of GMM data into two sets of normally distributed samples as a fitting model. Calculate the AIC value of this model and determine its fit to this set of data. We use several sets of Gaussian mixture models in Fig. \ref{2}, each with 1000 points selected, to calculate the AIC values under different threshold models. And compare it with the EM algorithm, a commonly used method for estimating Gaussian mixture models.

\begin{table}[htbp]
	\caption{NON-GAUSSIAN NOISES}
	\centering
	\label{tab:1}
	\resizebox{\linewidth}{!}{
		\begin{tabular}{cccccc}
			\hline
			Noise Distribution & $AI{C_{0.5\sigma }}$ & $AI{C_{0.8\sigma }}$ & $AI{C_\sigma }$ &  $AI{C_{1.2\sigma }}$ & $AI{C_{EM }}$ \\ \hline
			$0.9 N\left(0,0.1^{2}\right)+0.1 N\left(0,5^{2}\right)$ & -797 & -153 & -808 & -954 & 2043 \\ \hline
			$0.9 N\left(0,0.5^{2}\right)+0.1 N\left(0,10^{2}\right)$ & 1997 & 2072 & 2144 & 2426 & 2547 \\ \hline
			$0.8 N\left(0,0.1^{2}\right)+0.2 N\left(0,5^{2}\right)$ & 1249 & 5007 & 9186 & 14922 & 875 \\ \hline
			$0.8 N\left(0,0.5^{2}\right)+0.2 N\left(0,10^{2}\right)$ & 2657 & 3123 & 3792 & 5202 & 3380 \\ \hline
	\end{tabular}}
\end{table}

According to the results shown in the table, when a Gaussian mixture model consists of a component with a large variance and a small proportion, and another component with a small variance and a large proportion, the estimation performance of the EM algorithm is weaker than that of the thresholding method. This Gaussian mixture model scenario corresponds to background pulse noise. However, using EM algorithm for iterative computations on each point is extremely complex and cumbersome. On the other hand, the thresholding method is simple and efficient, making it suitable for online estimation.In Table \ref{tab:1}, the best fit occurs when the threshold falls between $0.5\sigma$ and $\sigma$, serving as the basis for subsequent iterations.

\subsubsection{Distinguish Noise Types}
In a set of GMM models, the distribution can be represented by equation \ref{eqn 13}.
\begin{equation}
	\label{eqn 13}
	{{r}_{fuse}}\sim N\left( \sum\limits_{i=1}^{j}{{{\mu }_{i}}},\sigma _{fuse}^{2} \right)
\end{equation}

Because the mean value of noise is usually 0, the fused data follows

\begin{equation}
	{{r}_{fuse}}\sim N\left( 0,\sigma _{fuse}^{2} \right)
\end{equation}
Now, we get the threshold value $\sigma _{fuse}$.

We take this threshold value as the basis for judging whether the current iterative observation noise belongs to impulse noise. If the difference of observed value and prior value is greater than the threshold value, we think this is a pulse noise.

\subsection{Proposed T-VBKF}
\begin{equation} 
	x_{k}={{A}_{k-1}}{{x}_{k-1}}+{q}_{k-1}
	\label{28}
\end{equation}
\begin{equation}
	y_{k}^{}={{H}_{k}}{{x}_{k}}+r_{k}^{i}
\end{equation}
\begin{equation}
	\begin{aligned}	
		\alpha _{k,m}^ -  = {\rho _m}{\alpha _{k - 1,m}},\quad m = 1, \ldots ,d\\
		\beta _{k,m}^ -  = {\rho _m}{\beta _{k - 1,m}},\quad m = 1, \ldots ,d
	\end{aligned}
\end{equation}
\begin{equation}
	\sigma _{k,m}^{{i}}=\frac{{{\beta }_{k,m}^{i}}}{{{\alpha }_{k,m}^{i}}} 
\end{equation}
\begin{equation}
	\Sigma _{k}^{i}=diag\left( {{(\sigma _{k,1}^{i})}^{2}},...,{{(\sigma _{k,d}^{i})}^{2}} \right)
\end{equation}
where ${{r}_{i}}\sim N\left( {{\mu }_{i}},\sigma _{i}^{2} \right)$.

When the iteration starts, we first use the VBAKF algorithm to filter and estimate the data. At the end of this phase, we obtain a set of observations and a set of Measurement Residual predictions. Using these two sets of data, we can calculate the required threshold value. We can use this threshold to group estimate the iteration points in the next stage.

In subsequent synchronous estimation algorithms, thresholds are recalculated for each specific number of iteration points(such as each 100 times) to enhance the timeliness of the algorithm.

\begin{equation} 
	{(\beta _{k,m}^l)^{(n + 1)}} = {(\beta _{k,m}^l)^{(n)}},(l \ne i)
\end{equation}
\begin{equation}
	{K_k} = {P_{k|k - 1}}H_k^T{({H_k}{P_{k|k - 1}}H_k^T + \hat \Sigma _k^{(n)})^{ - 1}}
\end{equation}
\begin{equation}
	\hat x_k^{\left( {n + 1} \right)} = {\hat x_{k|k - 1}}{\rm{ + }}K_k\left( {{y_k} - {H_k}{{\hat x}_{k|k - 1}}} \right)
\end{equation}
\begin{equation}
	P_k^{\left( {n + 1} \right)} = {P_{k|k - 1}} - K_k{H_k}{P_{k|k - 1}}
\end{equation}
\begin{equation}
	\label{eqn 25}
	\begin{split}
		{(\beta _{k,m}^i)^{\left( {n + 1} \right)}} = {(\beta _{k,m}^i)^ - } + \frac{1}{2}\left( {{y_k} - {H_k}\hat x_k^{\left( {n + 1} \right)}} \right)_m^2\\
		+ \frac{1}{2}{\left( {{H_k}P_k^{\left( {n + 1} \right)}H_k^T} \right)_{mm}}
	\end{split}
\end{equation}

We use the T-VBKF algorithm to process the data. The parameters($\alpha$,$\beta$) that do not belong to the scope of Measurement Residual prediction of the current iteration remain the same as the results of the previous iteration.
\begin{equation}
	{{d}_{k}}=\left| {{y}_{k}}-{{H}_{k}}{{x}_{k}} \right|
\end{equation}

where ${d_k} \in \left[ {0,{\sigma _{fuse}}} \right)or\left[ {{\sigma _{fuse}}, + \infty } \right)$.

Use a threshold value to distinguish the current iteration point, when the error of the Measurement Residual prediction belongs to the $[0,{\sigma _{fuse}})$ interval, ……

And when the error of the Measurement Residual prediction belongs to the $[{\sigma _{fuse}},\infty )$ interval, ……

When the iteration point goes through the next specific number of iteration points, we update the threshold value
\begin{equation}
	\label{eqn 27}
	d' = c*{d^{new}} + (1 - c)*{d^{old}}
\end{equation}

where $c \in (0,1)$.

The update steps of T-VBKF are essentially limited to equation \ref{eqn 25} and \ref{eqn 27}, both are fixed point algorithms. The fixed point algorithm is one of the natural gradient descent algorithms, which has optimal convergence \cite{2009Recursive}.

\section{ILLUSTRATIVE EXAMPLE}
In this section, a linear dynamic model is described to demonstrate the results proposed in this paper and to analyze the performances of several Kalman filters, namely, the KF, VBAKF and T-VBKF.
\begin{equation}\label{eqn-38} 
	\begin{array}{l}
		{x_{\rm{k}}} = {A_{\rm{k}}}{x_{{\rm{k}} - 1}} + {B_{\rm{k}}}u + {w_{\rm{k}}}\\
		{y_k} = {H_k}{x_k} + {v_k}
	\end{array}
\end{equation}

\subsection{CV Model in Target Tracking}
\[{x_k} = \left( {\begin{array}{*{20}{c}}
		1&0\\
		0&1
\end{array}} \right){x_{k - 1}} + \left( {\begin{array}{*{20}{c}}
		1&0\\
		0&1
\end{array}} \right)u + {w_k}\]
\[{y_k} = \left( {\begin{array}{*{20}{c}}
		1&0\\
		0&1
\end{array}} \right){x_k} + {v_k^{fuse}}\]
where $w_k\sim N\left( {0},Q \right)$ and 
$Q = \left( {\begin{array}{*{20}{c}}
		{{{0.1}^2}}&0\\
		0&{{{0.1}^2}}
\end{array}} \right)$.

We consider using Gaussian mixture noise as the observation noise and we consider a constant velocity (CV) model for two-dimensional target tracking under two types of noise\cite{6}.
\begin{equation}\label{eqn-39} 
	{\mathop{\rm RMSE}\nolimits} (i) = \sqrt {\frac{1}{M}\sum\limits_{j = 1}^M {\left( {{{\left( {x_{i,1}^j - \hat x_{i,1}^j} \right)}^2} + {{\left( {x_{i,2}^j - \hat x_{i,2}^j} \right)}^2}} \right)} }
\end{equation}

\subsubsection{GMM to Approximate Common Non-Gaussian Noise}

\begin{figure}[htb]
	\begin{minipage}[b]{1.0\linewidth}
		\centering
		\includegraphics[scale=0.8]{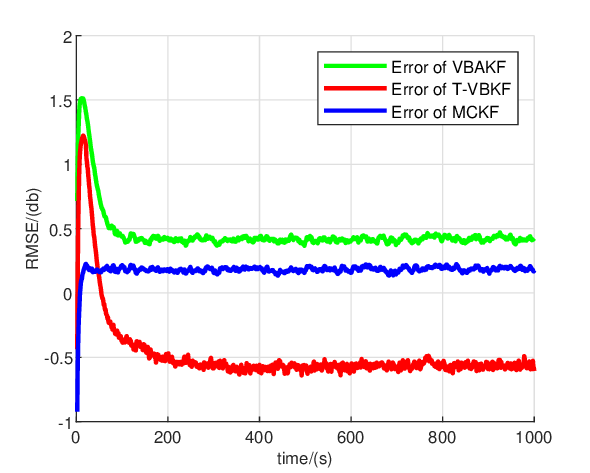}
	\end{minipage}
	\captionsetup{skip=0pt}
	\caption{RMSE of VBAKF, MCKF, T-VBKF under GMM\\(${r}\sim (0.8 N\left(0,0.5^{2}\right)+0.2 N\left(0,10^{2}\right))$)}
	\label{key3}
\end{figure}

We conducted 1000 comparative experiments, selecting 1000 points for each experiment. As shown in the Fig.\ref{key3} , compared with MCKF, we can clearly see that T-VBKF performs better than VBAKF and MCKF in the middle and later stages of iteration, with a faster descent rate than VBAKF. Due to the inability of MCKF to perform online estimation, the rate of error reduction is almost non-existent. T-VBKF can perform online real-time estimation, so there is a significant error in the initial stage\cite{5}.
\begin{table}[htbp]
	\caption{RMSES UNDER DIFFERENT GMM}
	\centering
	\label{tab 2}
	\resizebox{\linewidth}{!}{
		\begin{tabular}{cccccc}
			\hline
			&Noise Distribution&VBAKF&MCKF&T-VBKF&\\
			\hline
			&$0.8 N\left(0,0.1^{2}\right)+0.2 N\left(0,10^{2}\right)$&1.6187&1.1839&0.4218&\\
			&$0.8 N\left(0,0.1^{2}\right)+0.2 N\left(0,5^{2}\right)$&1.0604&0.85749&0.68705&\\
			&$0.8 N\left(0,0.5^{2}\right)+0.2 N\left(0,10^{2}\right)$&1.5999&1.1945&0.6621&\\
			&$0.9 N\left(0,0.1^{2}\right)+0.1 N\left(0,10^{2}\right)$&1.3612&0.9110&0.3274&\\
			&$0.9 N\left(0,0.1^{2}\right)+0.1 N\left(0,5^{2}\right)$&0.8573&0.6392&0.2812&\\
			&$0.9 N\left(0,0.5^{2}\right)+0.1 N\left(0,10^{2}\right)$&1.3643&0.9334&0.6621&\\
			\hline
	\end{tabular}}
\end{table}

To demonstrate the improvement of T-VBKF, we employed various non-Gaussian noises. In each experiment, we conducted 100 iterations, each comprising 1000 data points. We calculated the RMSE value for each point, summed them for the 1000 iterations, and then averaged across 100 experiments. The results are presented in the table below Table \ref{tab 2}. According to the table, the proposed T-VBKF algorithm exhibits better performance when confronted with strong impulsive noise\cite{2018Switching}.

\section{Conclusion}
This paper proposes an algorithm based on variational Bayesian Kalman filtering to process Gaussian mixture models, and tests and evaluates its performance through Matlab simulation. The results show that the algorithm can effectively process Gaussian mixture models, and has good robustness and computational efficiency. This algorithm exhibits good performance in processing noisy data and can play an important role in practical applications. In the future, we will further improve the algorithm and explore its applications in other fields.

\begin{spacing}{1}
	\bibliographystyle{IEEEbib}
	\bibliography{strings,refs}
\end{spacing}

\end{document}